\RequirePackage{fixltx2e}
\documentclass[a4paper]{isp}
\usepackage[american]{babel}
\usepackage{graphicx}
\usepackage{amsmath,amsthm,mathtools}
\usepackage{url}
\usepackage[capitalise,nameinlink]{cleveref}
\usepackage{xcolor}
\def\aap{A\&A\,  }

\def\aj{AJ  }
\def\apj{ApJ\,  }
\def\apjl{ApJ Letters,  }
\def\araa{ARA\&A  }

\def\cjaa{Chinese Astronomy and Astrophysics}

\def\mnras{MNRAS\,  }

\def\sovast{Soviet Astronomy} 
 
\def\snr1993j{SN\,1993J~}
\def\sigmad{$\Sigma-D~$}
\begin{document}
\pdfgentounicode=1
\title
{
Energy Conservation in the thin-layer approximation:
V. The surface brightness in supernova remnants  
}
\author{Lorenzo Zaninetti}
\institute{
Physics Department,
 via P. Giuria 1, I-10125 Turin, Italy \\
 \email{zaninetti@ph.unito.it}
}

\maketitle

\begin {abstract}
Two new equations of motion for a 
supernova remnant  (SNR) are derived in the framework 
of energy conservation for the thin-layer approximation.
The first one is based on an inverse square law
for the surrounding density  and the 
second one  on a  non-cubic dependence 
of the swept mass.
Under the assumption   that  the observed radio-flux  scales  
as the flux of kinetic energy,   two  scaling laws 
are derived for the temporal evolution of  
the  surface brightness of  SNRs.
The  astrophysical applications  cover two galactic  
samples of surface brightness and an extragalactic one.
\end{abstract}

\section{Introduction}

The  surface brightness  versus diameter, (\sigmad),  for  
supernova remnants (SNRs)
were initially analysed  from a theoretical 
point of view in the framework of the
initial conditions  for the time evolution 
of  SNRs \cite{Shklovskii1960,vanderLaan1962,Gull1973,Caswell1979}.
Some generic information on \sigmad can be found
in  reviews of  SNRs \cite{Woltjer1972,Long2017}.

The astrophysical approach to the \sigmad  
has always  mixed the observations with 
the theory and the statistics.
We select some items among others:
a catalogue of 25 SNRs 
has been compiled by \cite{Poveda1968}
with the conclusion that 
SNRs in the galactic halo have diameters greater than 
those in the
disk,
the evolutionary properties of 
SNRs have been deduced from  
observations with the Molonglo and Parkes radio telescopes 
\cite{Clark1976},
the  \sigmad relationship was used 
to fix the scale for distances of SNRs \cite{Huang1985},
the diameters, luminosities, surface brightness, galactic heights
for  231 SNRs  were processed in the
framework of the Sedov solution \cite{Xu2005},
a  new high-latitude SNR \cite{Kothes2017} was analysed,
an updated radio \sigmad relationship was derived \cite{Vukotic2019}
and
the phases of some supernova remnant have been determined  from the 
observations  \cite{Hu2019}.
In order to derive  the \sigmad relationship 
some hypotheses  on the single  equation of motion 
should be made, e.g., the conservation
of the momentum \cite{Zaninetti2012a}.
Here conversely we will apply the energy conservation 
in the framework of the thin-layer approximation
and we will  derive an analytical expression for the 
\sigmad relationship.
This paper 
derives two new equations
of motion in the framework  of 
energy conservation for the thin-layer 
approximation, see Section \ref{section_motion},
applies the two  analytical  expressions 
for the  surface brightness to two
galactic samples, see Section \ref{section_galactic},
and  to an 
extragalactic catalog, 
see Section \ref{section_extragalactic}.

\section{The equations of motion}

\label{section_motion}

The conservation of kinetic energy in
spherical coordinates
in  the framework of the thin-layer approximation,
here taken to be an assumption,  
states that  
\begin{equation}
\frac{1}{2} M_0(r_0) \,v_0^2 = \frac{1}{2}M(r) \,v^2 
\quad ,
\label{cons_rel_energy}
\end{equation}
where $M_0(r_0)$ and $M(r)$ are the swept masses at $r_0$ and $r$,
while $v_0$ and $v$ are the velocities of the thin layer at $r_0$ and $r$.
We now present two equations of motion for SNRs 
and the back-reaction for one of the two.

\subsection{The inverse square law}

The medium 
around the SN
is assumed to scale as an inverse square law
\begin{equation}
 \rho (r;r_0)  = \{ \begin{array}{ll}
            \rho_c                      & \mbox {if $r \leq r_0 $ } \\
            \rho_c (\frac{r_0}{r})^2    & \mbox {if $r >     r_0 $.}
            \end{array}
\label{piecewiseinvsquare},
\end{equation}
where 
$\rho_c$ is the density at $r=0$
and  
$r_0$ is the radius after which the density 
starts to decrease.
When  
the conservation of energy is applied 
the velocity as a function of the radius is  
\begin{equation}
v(r;r_0,v_0)=
-{\frac {\sqrt {- \left( 2\,r_{{0}}-3\,r \right) r_{{0}}}v_{{0}}}{2\,r
_{{0}}-3\,r}}
\quad  .
\end{equation}
The trajectory, i.e. the radius as a function of time,  is
\begin{equation}
r(t;t_0,r_0,v_0)= 
\frac{1}{6}\,\sqrt [3]{2}\sqrt [3]{r_{{0}}} \left(  \left( 9\,t-9\,t_{{0}}
 \right) v_{{0}}+2\,r_{{0}} \right) ^{2/3}+\frac{2}{3}\,r_{{0}}
\quad,
\label{rtinversesquare}
\end{equation}
and the velocity as a function of 
 time is 
\begin{equation} 
v(t;t_0,r_0,v_0) =
{\frac {\sqrt [3]{2}\sqrt [3]{r_{{0}}}v_{{0}}}{\sqrt [3]{ \left( 9\,t-
9\,t_{{0}} \right) v_{{0}}+2\,r_{{0}}}}}
\quad .
\end{equation}
More details can be found  in \cite{Zaninetti2020a}.
The   
rate of transfer of
mechanical energy, $L_{m}$, is
\begin{equation}
L_{m}(t) = \frac{1}{2}\rho (t)4 \pi r(t)^2 v(t)^3
\quad ,
\end{equation}
where $\rho(t)$,$r(t)$ and  $v(t)$ are the instantaneous
density, radius and velocity of the SN.
We now assume     that the density in front of the
advancing  expansion scales as
\begin{equation}
\rho(t) = \rho_0 ( \frac{r_0}{r(t)} ) ^{d}
\quad ,
\end{equation}
where $r_0$ is the radius at $t_0$ 
and   $d$ is a parameter  which allows
matching the observations.
The mechanical luminosity is now 
\begin{equation}
L_{m}(t) = \frac{1}{2}   
\rho_0 ( \frac{r_0}{r(t)} ) ^{d}
 4 \pi r(t)^2 v(t)^3
\quad .
\label{eqnlmtrho}
\end{equation}
In the case here analysed of the  inverse square profile
for density we have
\begin{equation}
L_{m}(t;v_0,t_0,r_0)=\frac{DLM}{81\,v_{{0}}t-81\,v_{{0}}t_{{0}}+18\,r_{{0}}}
\quad ,
\end{equation}
where
\begin{eqnarray}
DLM=\rho_{{0}}{6}^{d} \left( {r_{{0}} \left( \sqrt [3]{2}\sqrt [3]{r_{{0}}
} \left(  \left( 9\,t-9\,t_{{0}} \right) v_{{0}}+2\,r_{{0}} \right) ^{
{\frac{2}{3}}}+4\,r_{{0}} \right) ^{-1}} \right) ^{d}
\nonumber \\
\times
\pi\, \left( 
\sqrt [3]{2}\sqrt [3]{r_{{0}}} \left(  \left( 9\,t-9\,t_{{0}} \right) 
v_{{0}}+2\,r_{{0}} \right) ^{{\frac{2}{3}}}+4\,r_{{0}} \right) ^{2}r_{
{0}}{v_{{0}}}^{3}
\quad  .
\end{eqnarray} 
We now assume  that 
the observed luminosity, $L_{\nu}$,
in a given band
denoted by the frequency  $\nu$ is  proportional
to the mechanical  luminosity
\begin{equation}
L_{\nu}(t)  = cost * L_{m}(t)  
\quad ,
\end{equation}
where  $L_{\nu}$ is the observed
radio luminosity in a given band
and $cost$ a constant 
which resolves the  mismatch between  theory and observations.
The surface brightness
is the luminosity divided by  the 
interested area 
\begin{equation}
\Sigma = \frac{L_{\nu}(t)}{\pi r(t)^2}
\label{sigmadef}  
\quad ,
\end{equation} 
which is
\begin{equation}
\Sigma(t;v_0,t_0,r_0) ={\it cost}\frac
{
4\,\,\rho_{{0}}{6}^{d} \left( {\frac {r_{{0}}}{\sqrt [3]{2}
\sqrt [3]{r_{{0}}} \left(  \left( 9\,t-9\,t_{{0}} \right) v_{{0}}+2\,r
_{{0}} \right) ^{2/3}+4\,r_{{0}}}} \right) ^{d}r_{{0}}{v_{{0}}}^{3}
}
{
9\,v_{{0}}t-9\,v_{{0}}t_{{0}}+2\,r_{{0}}
}
\quad .
\label{sigmadtinversesquare}
\end{equation}

\subsection{Non cubic dependence}

\begin{figure*}
\begin{center}
\includegraphics[width=7cm]{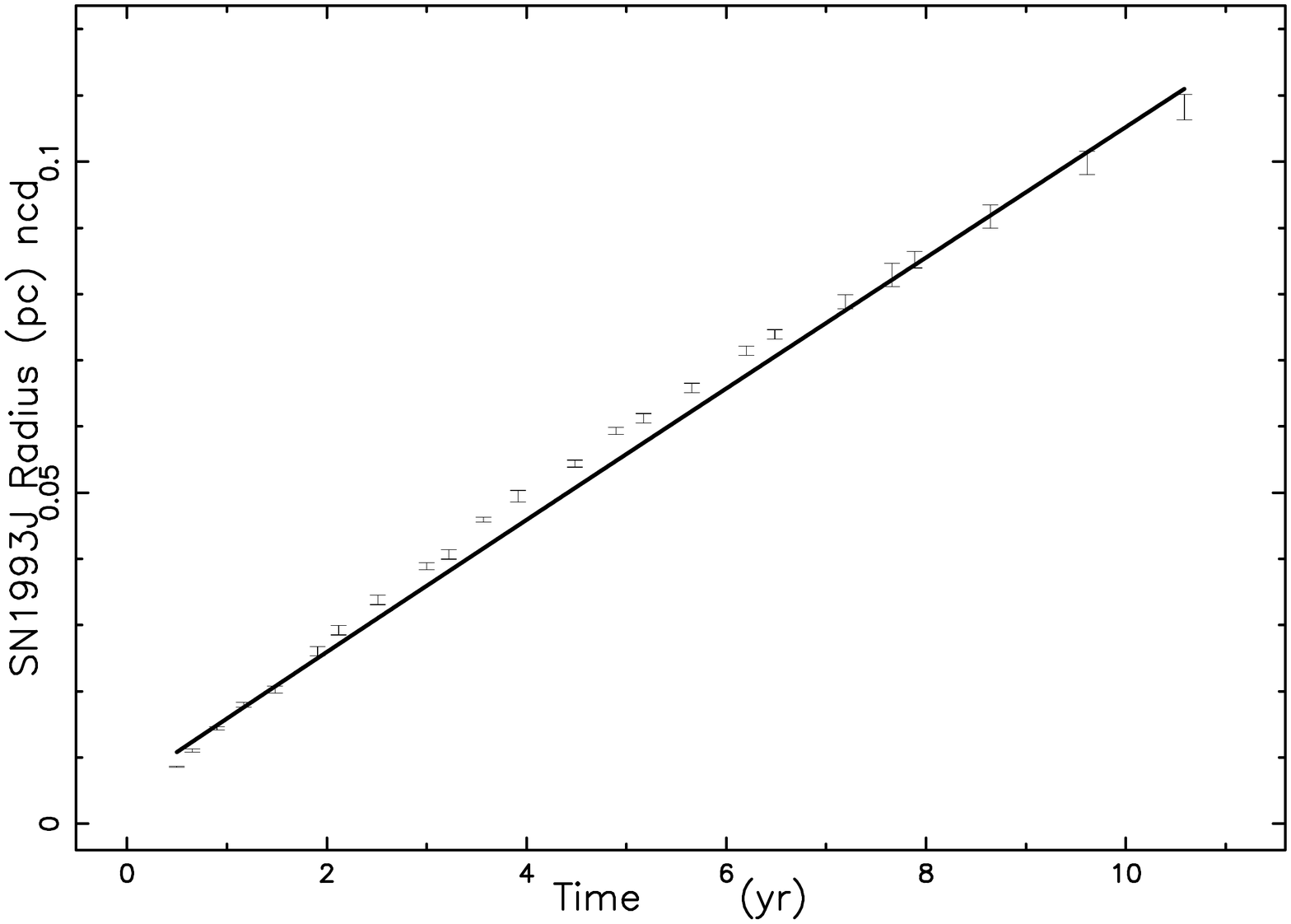}
\end {center}
\caption
{
Analytical solution 
for the NCD case with parameters 
$r_0=0.006\,pc$, $v_0 =10000\,km/s$, $t_0=0.026\,yr$ and
$\delta =0.2$, which gives $\chi^2=3751$. 
The  astronomical data of \snr1993j
are represented with vertical error bars.
}
\label{ncd_sn1993j}
    \end{figure*}

The swept  mass  is assumed to scale as   
\begin{equation}
 M(r;r_0,\delta)  = \Bigg \{ \begin{array}{ll}
M_0                            & \mbox {if $r \leq r_0 $ } \\
M_0 (\frac{r}{r_0})^{\delta}   & \mbox {if $r >     r_0 $}
            \end{array}
\label{piecewisencd},
\end{equation}
where 
$M_0$ is the swept mass at  $r=r_0$,
and $\delta$ is a regulating parameter less than 3.
The  differential equation  of the first order 
which regulates
the motion  is obtained by inserting the above $M(r)$ 
in equation (\ref{cons_rel_energy})
\begin{equation}
\frac{dr(t;r_0,v_0,\delta)}{dt}=
\frac
{
v_{{0}}
}
{
\sqrt {{r}^{\delta}{r_{{0}}}^{-\delta}}
}
\quad ,
\label{diffeqnncd}
\end{equation}
which has as the solution
\begin{equation}
r(t;r_0,v_0,\delta) 
=
\exp{\Bigg (\frac{ER}{\delta+2}\Bigg )}
\label{rt_ncd}
\end{equation}
where
\begin{eqnarray}
ER=
\ln  \Big ( {r_{{0}}}^{\delta+1}\delta\,tv_{{0}}-{r_{{0}}}^{\delta+1}
\delta\,t_{{0}}v_{{0}}+2\,{r_{{0}}}^{\delta+1}tv_{{0}}
\nonumber \\
-2\,{r_{{0}}}^{
\delta+1}t_{{0}}v_{{0}}+{r_{{0}}}^{\delta+2}+{\frac {{r_{{0}}}^{\delta
}{\delta}^{2}{t}^{2}{v_{{0}}}^{2}}{4}}-{\frac {{r_{{0}}}^{\delta}{
\delta}^{2}tt_{{0}}{v_{{0}}}^{2}}{2}}+{\frac {{r_{{0}}}^{\delta}{
\delta}^{2}{t_{{0}}}^{2}{v_{{0}}}^{2}}{4}}
\nonumber \\
+{r_{{0}}}^{\delta}\delta\,{
t}^{2}{v_{{0}}}^{2}-2\,{r_{{0}}}^{\delta}\delta\,tt_{{0}}{v_{{0}}}^{2}
+{r_{{0}}}^{\delta}\delta\,{t_{{0}}}^{2}{v_{{0}}}^{2}+{r_{{0}}}^{
\delta}{t}^{2}{v_{{0}}}^{2}-2\,{r_{{0}}}^{\delta}tt_{{0}}{v_{{0}}}^{2}
+{r_{{0}}}^{\delta}{t_{{0}}}^{2}{v_{{0}}}^{2} \Big) 
\quad .
\end{eqnarray}
The velocity is 
\begin{equation}
v(t;r_0,v_0,\delta) 
=
\frac{NV}
{
\left( \delta+2 \right)  \left( \delta\,tv_{{0}}-\delta\,t_{{0}}v_{{0
}}+2\,tv_{{0}}-2\,t_{{0}}v_{{0}}+2\,r_{{0}} \right) ^{2}
}
\label{vfirst}
\end{equation}
where
\begin{eqnarray}
NV =
v_{{0}} \left( {r_{{0}}}^{\delta} \left( v_{{0}} \left( \delta+2
 \right)  \left( t-t_{{0}} \right) +2\,r_{{0}} \right) ^{2} \right) ^{
 \left( \delta+2 \right) ^{-1}} 
\nonumber \\
\times \Big ( 
2\,{4}^{- \left( \delta+2
 \right) ^{-1}}{\delta}^{2}tv_{{0}}-2\,{4}^{- \left( \delta+2 \right) 
^{-1}}{\delta}^{2}t_{{0}}v_{{0}}+8\,{4}^{- \left( \delta+2 \right) ^{-
1}}\delta\,tv_{{0}}-8\,{4}^{- \left( \delta+2 \right) ^{-1}}\delta\,t_
{{0}}v_{{0}}+{4}^{{\frac {\delta+1}{\delta+2}}}\delta\,r_{{0}}
\nonumber  \\
+8\,{4}^
{- \left( \delta+2 \right) ^{-1}}tv_{{0}}-8\,{4}^{- \left( \delta+2
 \right) ^{-1}}t_{{0}}v_{{0}}+8\,{4}^{- \left( \delta+2 \right) ^{-1}}
r_{{0}} 
\Big )
\quad . 
\end{eqnarray} 
Figure \ref{ncd_sn1993j} 
reports the analytical solution
for the NCD case as given by equation (\ref{rt_ncd})
for \snr1993j.

The mechanical luminosity is  assumed to scale  as
in equation (\ref{eqnlmtrho}) 
and therefore
in the NCD case is  
\begin{eqnarray}
L_{m}(t;r_0,v_0,t_0,\delta)
=
\nonumber \\
16\,{\frac {{{\rm e}^{{\it EA}}}{v_{{0}}}^{3} \left( r_{{0}}{{\rm e}^{
{\it EB}}} \right) ^{d}\pi\,\rho_{{0}}}{ \left( \delta+2 \right) ^{3}
 \left( v_{{0}} \left( \delta+2 \right)  \left( t-t_{{0}} \right) +2\,
r_{{0}} \right) ^{6}}{S_{{2}}}^{3\, \left( \delta+2 \right) ^{-1}}
 \left( S_{{1}}{4}^{{\frac {\delta+1}{\delta+2}}}+{4}^{- \left( \delta
+2 \right) ^{-1}}v_{{0}}{\delta}^{2} \left( t-t_{{0}} \right) 
 \right) ^{3}}
\quad ,
\end{eqnarray}
where
\begin{eqnarray}
EA=
\frac{1}{\delta+2} 
\Big (
{
-4\,\ln    ( 2   ) +2\,\ln    ( 4\,v_{{0}}   ( \delta+2
   )    ( t-t_{{0}}   ) {r_{{0}}}^{\delta+1}+4\,{r_{{0}}}^{
\delta+2}+   (    ( {t}^{2}-2\,tt_{{0}}+{t_{{0}}}^{2}   ) {
\delta}^{2}
}
\nonumber  \\
{
+   ( 4\,{t}^{2}-8\,tt_{{0}}+4\,{t_{{0}}}^{2}   ) 
\delta+4\,{t}^{2}-8\,tt_{{0}}+4\,{t_{{0}}}^{2}   ) {v_{{0}}}^{2}{r
_{{0}}}^{\delta}   ) 
}
\Big )
\quad ,
\end{eqnarray}
and 
\begin{equation}
EB =
{\frac {2\,\ln  \left( 2 \right) -\ln  \left(  \left( v_{{0}} \left( 
\delta+2 \right)  \left( t-t_{{0}} \right) +2\,r_{{0}} \right) ^{2}{r_
{{0}}}^{\delta} \right) }{\delta+2}}
\quad ,
\end{equation}
\begin{equation}
S_1 =
 \left( \delta+1 \right)  \left( t-t_{{0}} \right) v_{{0}}+{\frac {r_{
{0}} \left( \delta+2 \right) }{2}}
\quad ,
\end{equation}
and 
\begin{equation}
S_2 =
 \left( v_{{0}} \left( \delta+2 \right)  \left( t-t_{{0}} \right) +2\,
r_{{0}} \right) ^{2}{r_{{0}}}^{\delta}
\quad .
\end{equation}
The surface brightness  is derived according to
equation (\ref{sigmadef}) and 
in the NCD case is  
\begin{eqnarray}
\Sigma (t;r_0,v_0,t_0,\delta) 
= 
\nonumber \\
{\it cost}
\frac
{
16\,\rho_{{0}} \left( {4}^{- \left( \delta+2 \right) ^{-1}}v_{{0}}{
\delta}^{2}t-{4}^{- \left( \delta+2 \right) ^{-1}}v_{{0}}{\delta}^{2}t
_{{0}}+{\it SC}\,{4}^{{\frac {\delta+1}{\delta+2}}} \right) ^{3}{
{\rm e}^{{\it SA}}}\,{v_{{0}}}^{3}{r_{{0}}}^{{\frac {
 \left( d+3 \right) \delta+2\,d}{\delta+2}}}{{\it SB}}^{{\frac {-6-6\,
\delta}{\delta+2}}}
}
{
 \left( \delta+2 \right) ^{3}
}
\quad ,
\label{sigmadtncd}
\end{eqnarray}
where
\begin{equation}
SA=  -{\frac {d \left( -2\,\ln  \left( 2 \right) +2\,\ln  \left( v_{{0}}
 \left( \delta+2 \right)  \left( t-t_{{0}} \right) +2\,r_{{0}}
 \right) +\delta\,\ln  \left( r_{{0}} \right)  \right) }{\delta+2}}
\quad ,
\end{equation}
\begin{equation}
SB=
v_{{0}} \left( \delta+2 \right)  \left( t-t_{{0}} \right) +2\,r_{{0}}
\end{equation}
and
\begin{equation}
SC=
\left( \delta+1 \right)  \left( t-t_{{0}} \right) v_{{0}}+{\frac {r_{
{0}} \left( \delta+2 \right) }{2}}
\quad .
\end{equation}

\begin{figure*}
\begin{center}
\includegraphics[width=7cm]{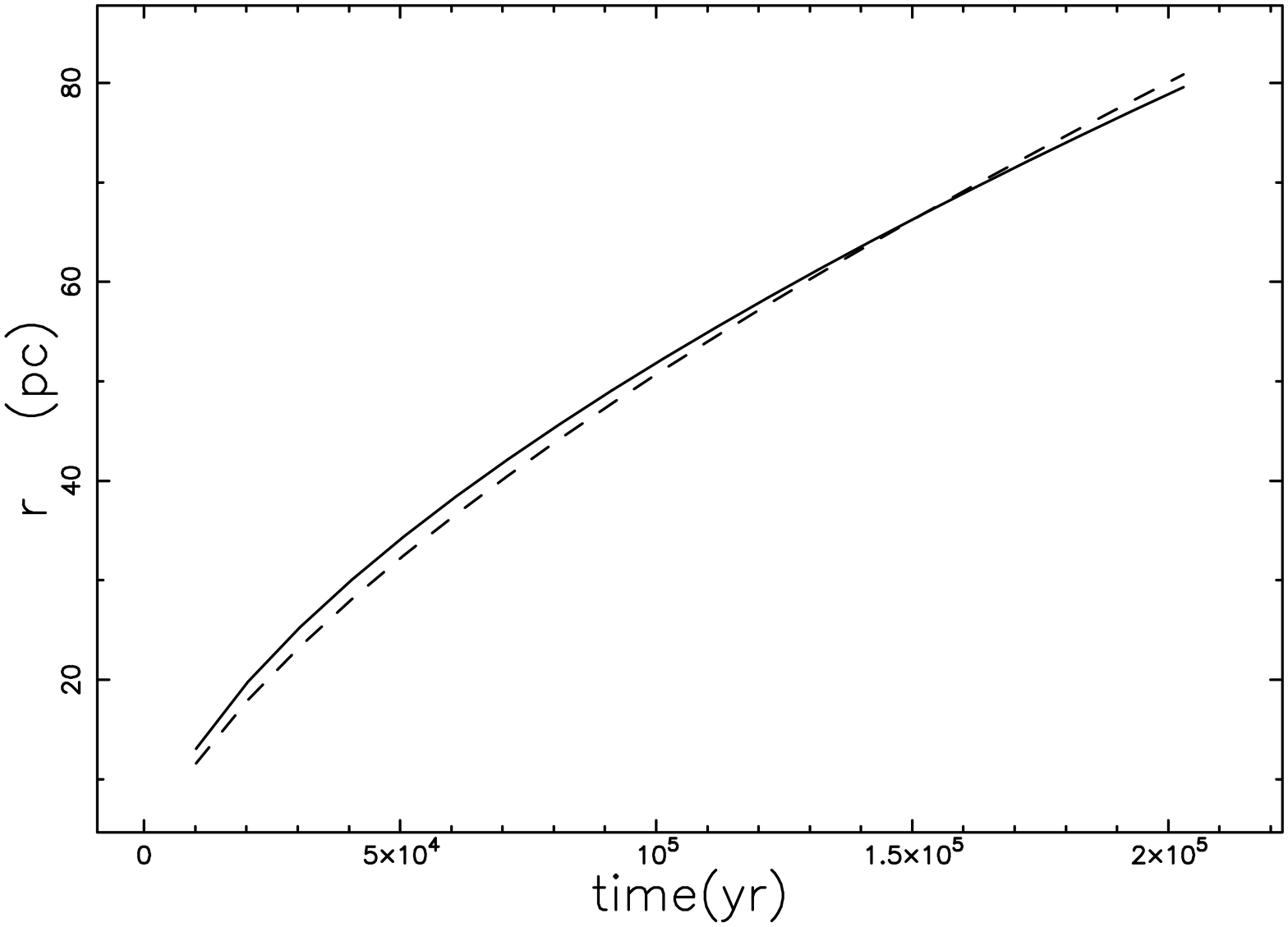}
\end {center}
\caption
{
Analytical solution 
for the inverse square  model (dashed line)  
when  
$r_0=1\,pc$,
$v_0=4000\,km/s$,
$t_0=10\,yr$,   
and  NCD model  (full line ) 
with the same parameters and
$\delta =1.3$.
}
\label{comparison}
    \end{figure*}

A comparison of the  two models 
here implemented for the trajectory
is reported in Figure \ref{comparison}.

\subsection{Non cubic dependence and back reaction}

The radiative losses per unit length  
are assumed to be  proportional to the   flux of momentum
as an assumption
\begin{equation}
- \epsilon \rho_s v^2 4\,\pi r^2 
\quad ,
\end{equation}
where $\epsilon$ is a constant and $rho_s$ is the density
in the thin advancing layer.

\begin{figure*}
\begin{center}
\includegraphics[width=7cm]{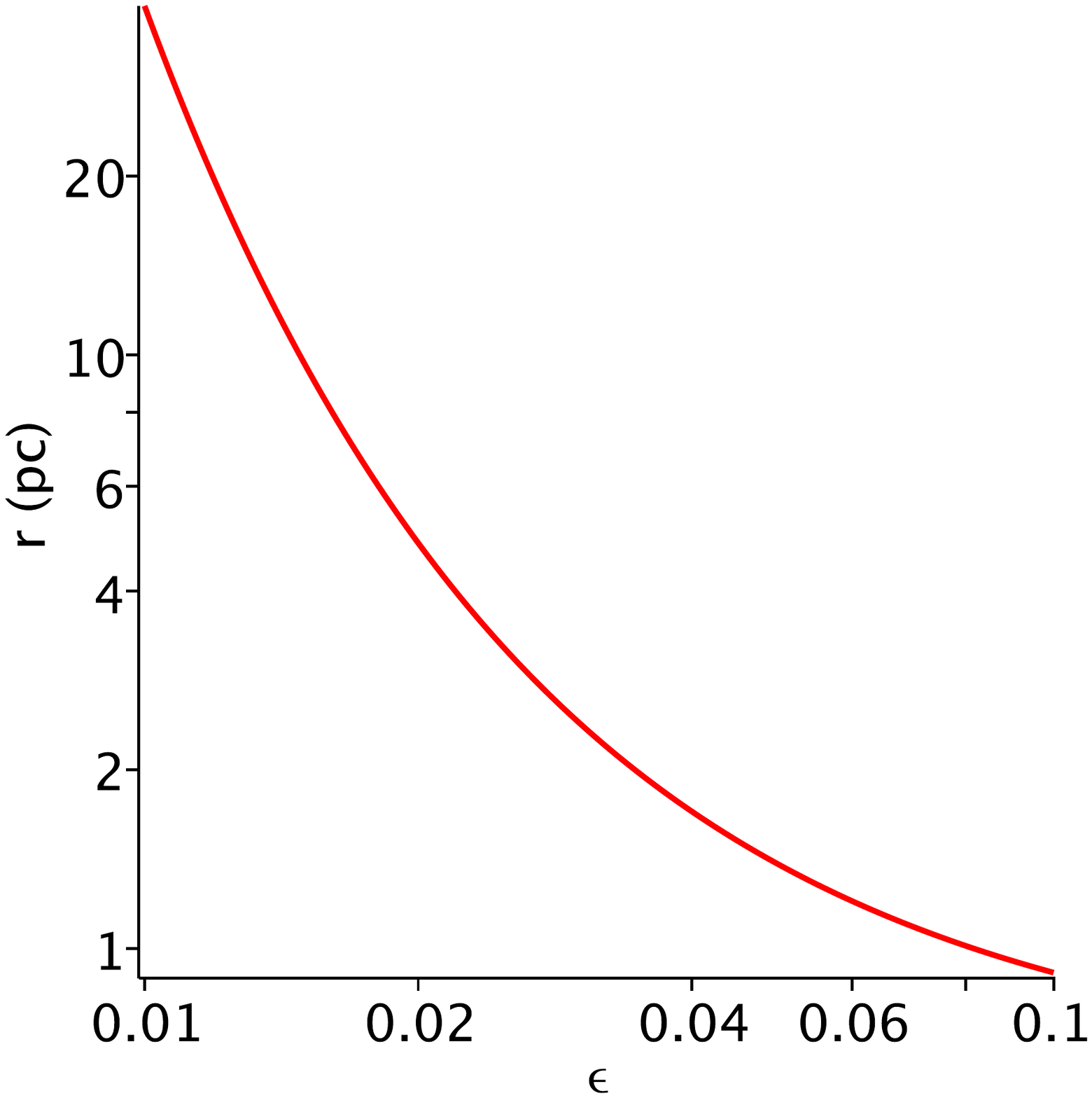}
\end {center}
\caption
{
Length of the SNR, $r_{back}$, 
when $r_0=0.6pc$ as function of $\epsilon$. 
}
\label{rncdback}
    \end{figure*}

The volume, $V$,  of the advancing layer 
is 
\begin{equation}
V = 4 \pi r^2 \Delta\,r
\end{equation}
with $\Delta r=r/12$,
therefore 
the above density for  the advancing layer
is
\begin{equation}
\rho_s=\frac{M(r;r_0,\delta)}{V}
\quad .
\end{equation}
Inserting in the above equation  the  velocity to the first order
as  given by equation~(\ref{vfirst}) 
the radiative losses, $Q(r;r_0,v_0,\delta,\epsilon)$, are
\begin{equation}
Q(r;r_0,v_0,\delta,\epsilon)=
-12\,{\frac {\epsilon\,M_{{0}}{v_{{0}}}^{2}}{r}}
\quad .
\label{lossesncdunit}
\end{equation}
The sum of the radiative  losses between $r_0$ and $r$ 
is given by the following integral, $L$,
\begin{equation}
L(r;r_0,v_0,\delta,\epsilon)=\int_{r_0}^r  Q(r;r_0,v_0,\delta,\epsilon) dr
=
-12\,\epsilon\,M_{{0}}{v_{{0}}}^{2}\ln  \left( r \right) +12\,\epsilon
\,M_{{0}}{v_{{0}}}^{2}\ln  \left( r_{{0}} \right) 
\quad .
\label{lossesncd}
\end{equation}
The  conservation of energy  in  the presence  
of  the back reaction due to the radiative losses
is 
\begin{equation}
{\frac {2\,\pi\,{r_{{0}}}^{3}{v}^{2}}{3} \left( {\frac {r}{r_{{0}}}}
 \right) ^{\delta}}+16\,\epsilon\,\pi\,{r_{{0}}}^{3}{v_{{0}}}^{2}\ln 
 \left( r \right) -16\,\epsilon\,\pi\,{r_{{0}}}^{3}{v_{{0}}}^{2}\ln 
 \left( r_{{0}} \right) ={\frac {2\,\pi\,{r_{{0}}}^{3}{v_{{0}}}^{2}}{3
}}
\quad . 
\label{eqnenergybackncd}
\end{equation}
An  analytical solution for the velocity to 
second order, $v_c(r;r_0,c_0,\delta,\epsilon)$, 
is
\begin{equation}
v_c(r;r_0,v_0,\delta,\epsilon)=
{r}^{-{\frac {\delta}{2}}}{r_{{0}}}^{{\frac {\delta}{2}}}\sqrt {24\,
\ln  \left( r_{{0}} \right) \epsilon-24\,\ln  \left( r \right) 
\epsilon+1}v_{{0}}
\label{vcorrected}
\quad .
\end{equation}
The  inclusion  of the back reaction  allows the evaluation of the 
SRS's maximum length, $r_{back}(r_0,\delta,\epsilon)$ ,  which can be derived 
by setting the above velocity equal to zero.
\begin{equation}
r_{back}(r_0,\delta,\epsilon) 
= 
{{\rm e}^{{\frac {24\,\ln  \left( r_{{0}} \right) \epsilon+1}{24\,
\epsilon}}}}
\quad .
\end{equation}
Figure \ref{rncdback} reports the 
finite radius of the advancing SNR as a function of $\epsilon$.

\section{Galactic   Application}

\label{section_galactic}

In the following  we  will process a  sample  of
data $x_i,y_i$ with $i$ varying between 1 and $N$
by a power law   fit  of the 
the  type 
\begin{equation}
y(x) = C\,x^{\alpha}
\quad  ,
\end{equation}
where $C$ and $\alpha$ are two constants to
be found from the sample.

\begin{figure*}
\begin{center}
\includegraphics[width=7cm]{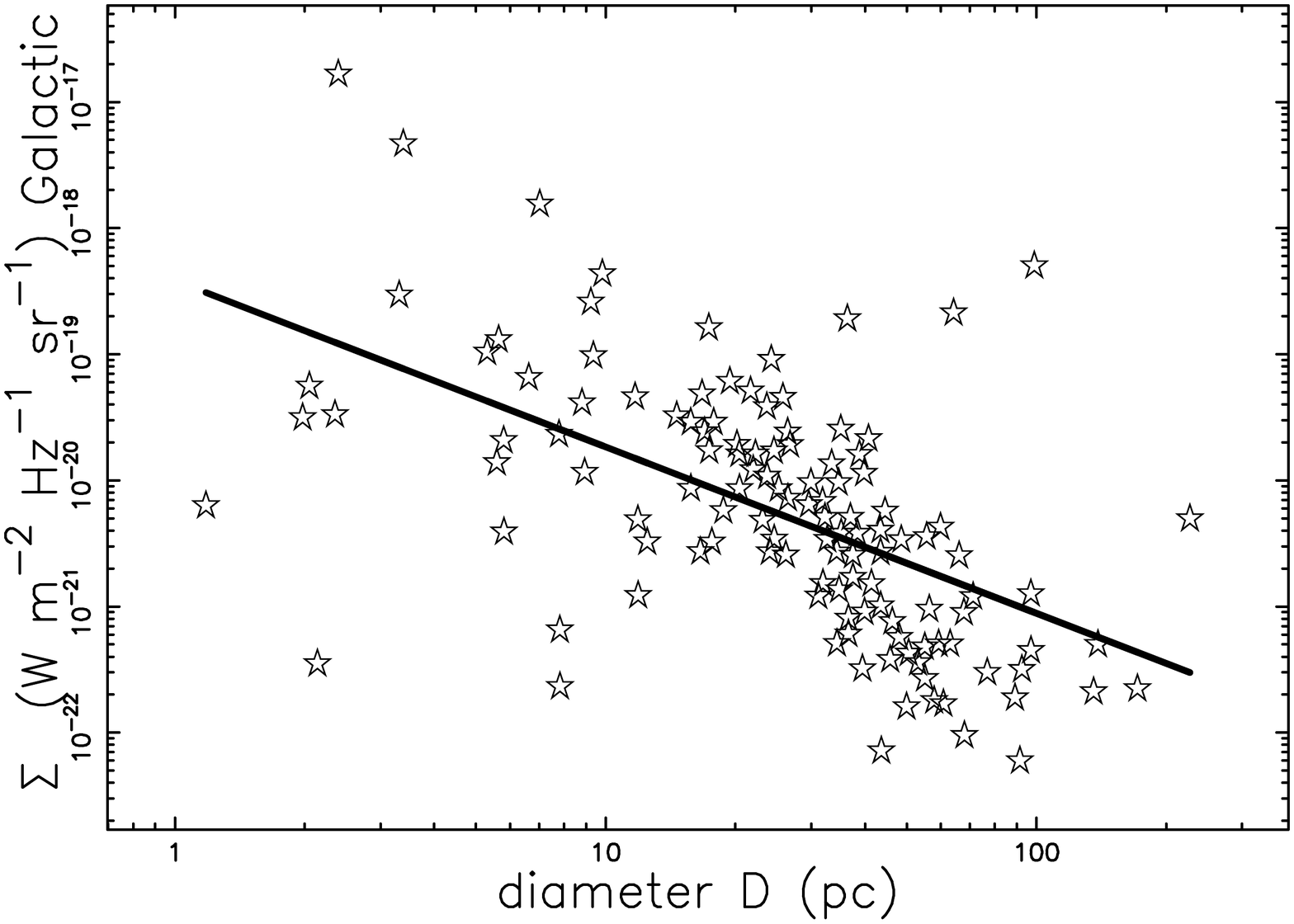}
\end {center}
\caption
{
Observed \sigmad relationship (empty stars)  
and  fitted  power law (full line)
with  parameters as in Table \ref{datafitsigmadgal}.
}
\label{galactic_sigmad}
    \end{figure*}

The  {\it first} source for   the  
observed  \sigmad relationship  
for the SNRs of our galaxy can be 
found  in Figure 3 of  \cite{Xu2005}
which  is  now digitized, see Figure \ref{galactic_sigmad}.

Table  \ref{datafitsigmadgal}    
reports 
the minimum  diameter, $D_{min}$,
the average diameter, $\overline{D}$,  
the maximum  diameter, $D_{max}$,  
the minimum   \sigmad, $\Sigma-D_{min}$,
the average  \sigmad, $\overline{\Sigma-D}$
and
the maximum   \sigmad, $\Sigma-D_{max}$
as well
as the two parameters of the power law fit.

\begin{table}
\caption
{
Statistics of the observed \sigmad galactic relationship
and the two parameters of the power law fit.
The theoretical parameters of the inverse square solution are
$r_0=1\,pc$,
$v_0=4000\,km/s$,
$d=1.1$,
$t_0=10\,yr$,   
$t_{min}= 2\,t_0$ 
and 
$t_{max}=2.05\,10^5$yr.
}
 \label{datafitsigmadgal}
 \[
 \begin{array}{lll}
 \hline
 \hline
 \noalign{\smallskip}
  parameter & observed   & theoretical \\
 \noalign{\smallskip}
 \hline
 \noalign{\smallskip}
D_{min} (pc)            & 1.18              & 3.44             \\
\overline{D}(pc)        & 37.17             & 37.4             \\
D_{max} (pc)            & 227.17            & 86.708           \\
\Sigma_{min} (pc)       & 5.96\,10^{-23}    &  3 \,10^{-20}    \\
\overline{\Sigma}(pc)   & 2.16\,10^{-19}    &  2.24\,10^{-19}  \\
\Sigma_{max} (pc)       & 1.66\,10^{-17}    &  2.54\,10^{-18}  \\
C                       & 3.83\,10^{-19}    &  1.32\,10^{-17}  \\
\alpha                  & -1.32             &  -1.33           \\
\noalign{\smallskip} 
\hline
 \end{array}
 \]
 \end {table}

A {\it second} source   for the \sigmad relationship 
is the  Green's catalog \cite{Green2019}
where the flux density in $Jy$  at $1\,GHz$, $S_1$,
and  the mayor and minor angular size in arcmin, $\theta$,
are reported for  295 SNRs.
The surface-brightness in SI is
\begin{equation}
\Sigma = 1.181\,10^{-19} \frac{S_1}{\theta}  
\frac{W}{m^2\, Hz \,sr} \quad SI
\quad ,
\end{equation}
but  after  \cite{Clark1976} 
the radio astronomers use the following 
conversion
\begin{equation}
\Sigma = 1.505\,10^{-19} \frac{S_1}{\theta}   
\frac{W}{m^2\, Hz \,sr} \quad astronomy
\quad ,
\end{equation}
which will also be adopted here.

\begin{figure*}
\begin{center}
\includegraphics[width=7cm]{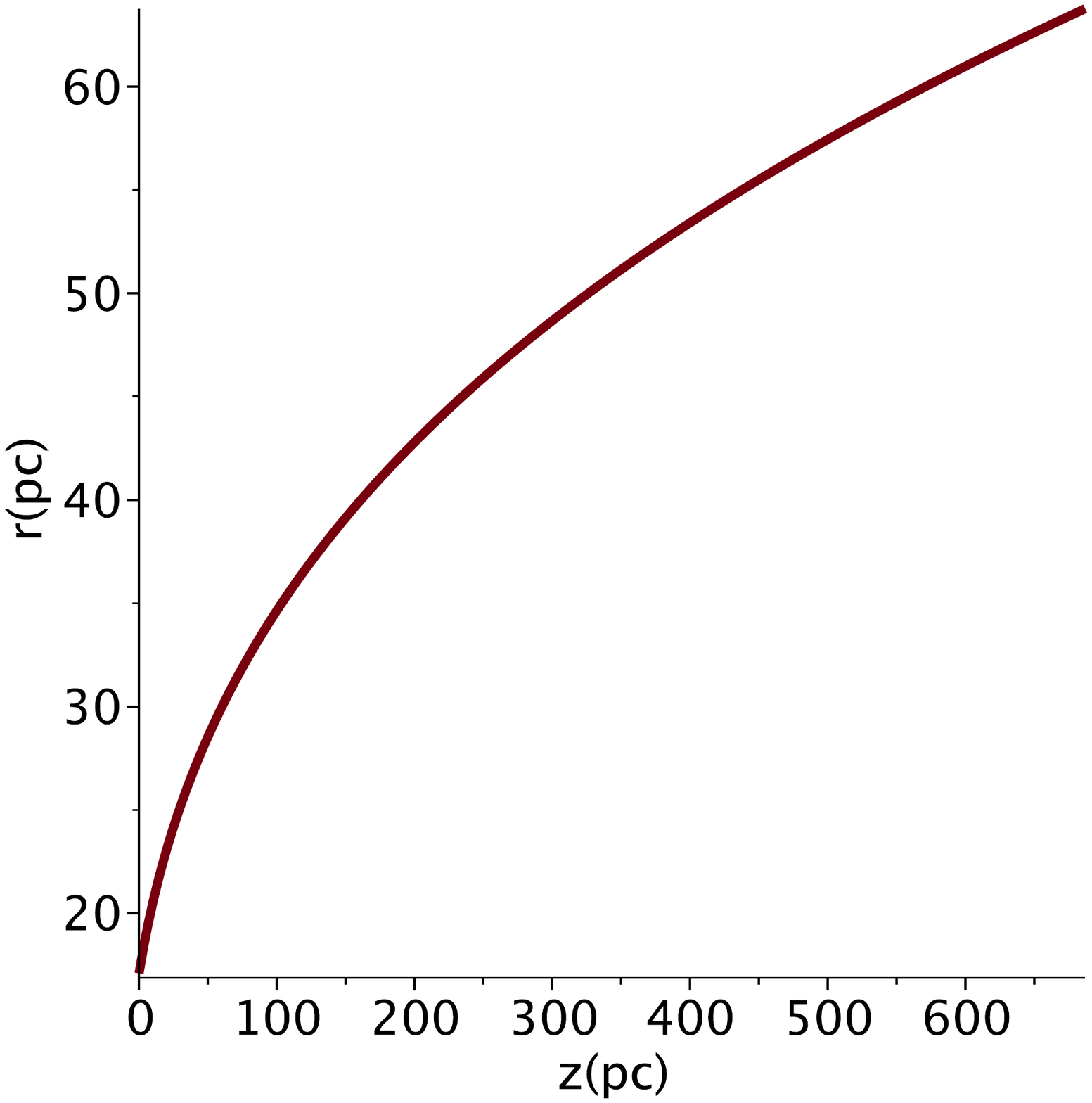}
\end {center}
\caption
{
Theoretical radius as given by the inverse square model,
see  Eq.~(\ref{rpczpc}), as  function of the galactic
height;
$v_0=4000 \,\frac{km}{s}$,
$t_0=10 \, yr$ 
and  $t=5\,10^4 yr$.
}
\label{rpczpc}
    \end{figure*}

The  probability   density function (PDF), $p(z)$,
to have an SNR as a function of  the galactic height $z$
is  characterized   by an exponential   PDF
\begin{equation}
p(z)  = \frac{1}{b} \exp {-\frac{z}{b}}
\quad  ,
\label{pdfsnrz}
\end{equation}
with  $b=83$ pc \cite{Xu2005}.
The linear  diameter ($D$) of an SNR 
increases   with  the galactic height see Figure 2 
in  \cite{Xu2005}
and in the framework of the model with an inverse square 
law model for density the parameter $r_0$ 
is chosen  to have the following dependence 
with the  
the galactic height
\begin{equation}
r_0 = 0.01 +0.5 (\frac{z}{z_{max}})
\label{condition1}
\quad ,
\end{equation}
where $z_{max}=687$ is the maximum   galactic height
which  belongs to  an SNR.
The above  assumption coupled  with an initial 
velocity for the inverse square 
law model for density  of  $v_0 = 4000 \frac{km}{s}$
for all the SNRs   will ensure a longer radius 
for SNRs with higher galactic heights,
see  Figure \ref{rpczpc}. 

We now simulate the   galactic  \sigmad relationship
for a number of theoretical SNRs, $N$, 
equal to that observed
according to the following rules
\begin{enumerate}
\item We randomly generate $N$   parameters $z$   
      according   to the exponential PDF (\ref{pdfsnrz})
\item At  each random value of $z$ we  
      associate  an initial parameter $r_0$ 
      according to the empirical equation~(\ref{condition1})
\item We randomly generate $N$ times, $t$, 
      according to the uniform distribution 
      between 
      a minimum value of time, $t_{min}$
      and
      a maximum value of time, $t_{max}$.
\item Given the parameters $r_0$, $v_0$, $t_0$ 
      and $t$ we evaluate the radius
      according to equation~(\ref{rtinversesquare}) 
      for an inverse power law  profile for density.
      The diameter is obtained doubling
      the above result. 
\item $\Sigma$ is now generated   according to 
equation~(\ref{sigmadtinversesquare})
      once  the regulating  parameter $d$  
      is provided
\end{enumerate}

\begin{figure*}
\begin{center}
\includegraphics[width=7cm]{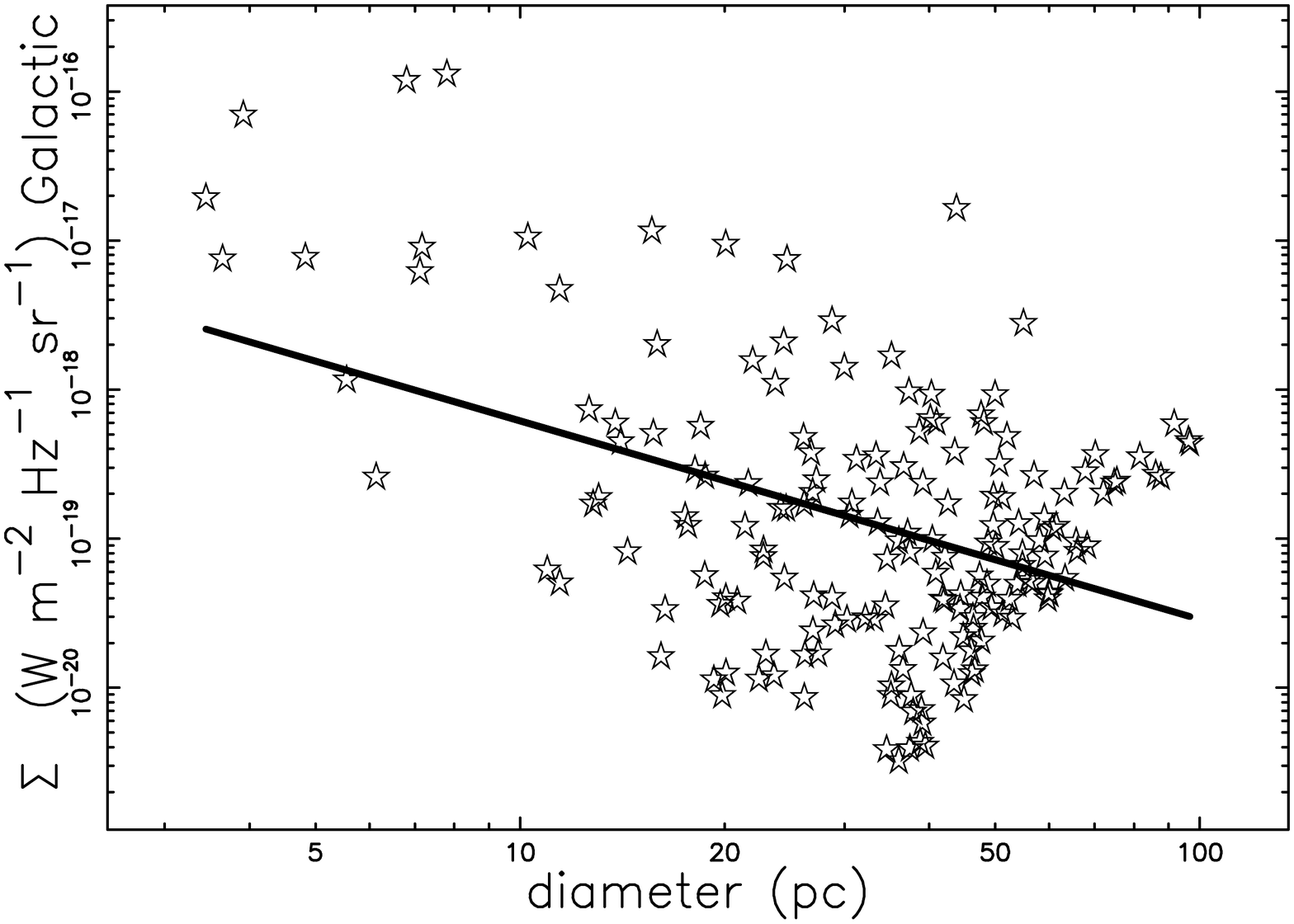}
\end {center}
\caption
{
Theoretical   \sigmad relationship (empty stars)  
and  fitted  power law (full line)
with  parameters as in Table \ref{datafitsigmadgal}.
}
\label{sigmad_galactic_theory}
    \end{figure*}

The  theoretical display  of the results
is reported in Figure \ref{sigmad_galactic_theory},
were we  matched   the three main parameters 
($\overline{D}(pc),\overline{\Sigma-D},\alpha $)
which for the  observations are  ($37.17, 2.16\,10^{-19}, -1.32$)
and   for our simulation are ($ 37.4, 2.24\,10^{-19}, -1.33$).

\begin{figure*}
\begin{center}
\includegraphics[width=7cm]{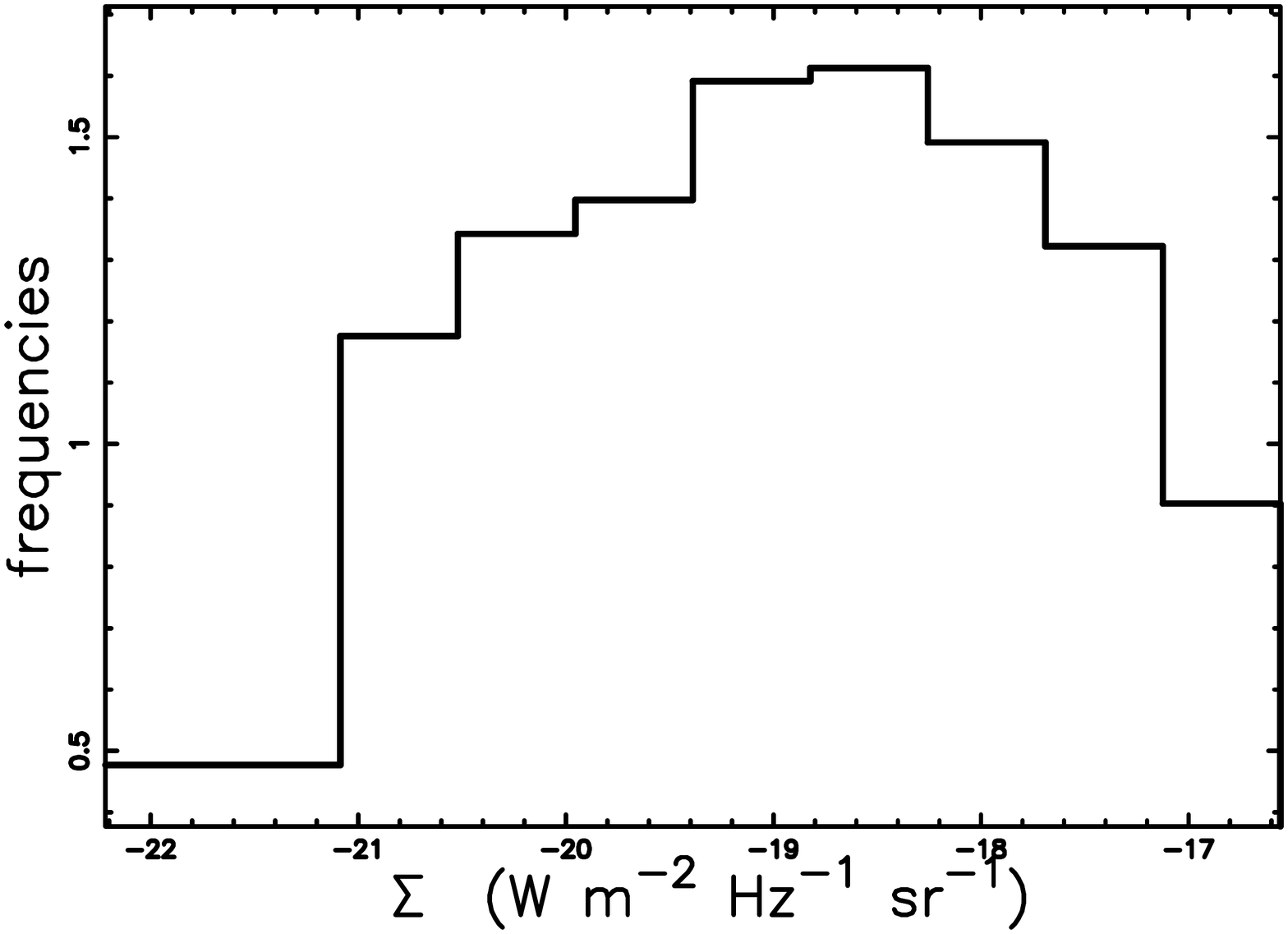}
\end {center}
\caption
{
Histogram  of the  \sigmad for 294 SNR  
as derived from the Green's catalog \cite{Green2019}.
}
\label{catalog_green}
    \end{figure*}

\begin{figure*}
\begin{center}
\includegraphics[width=7cm]{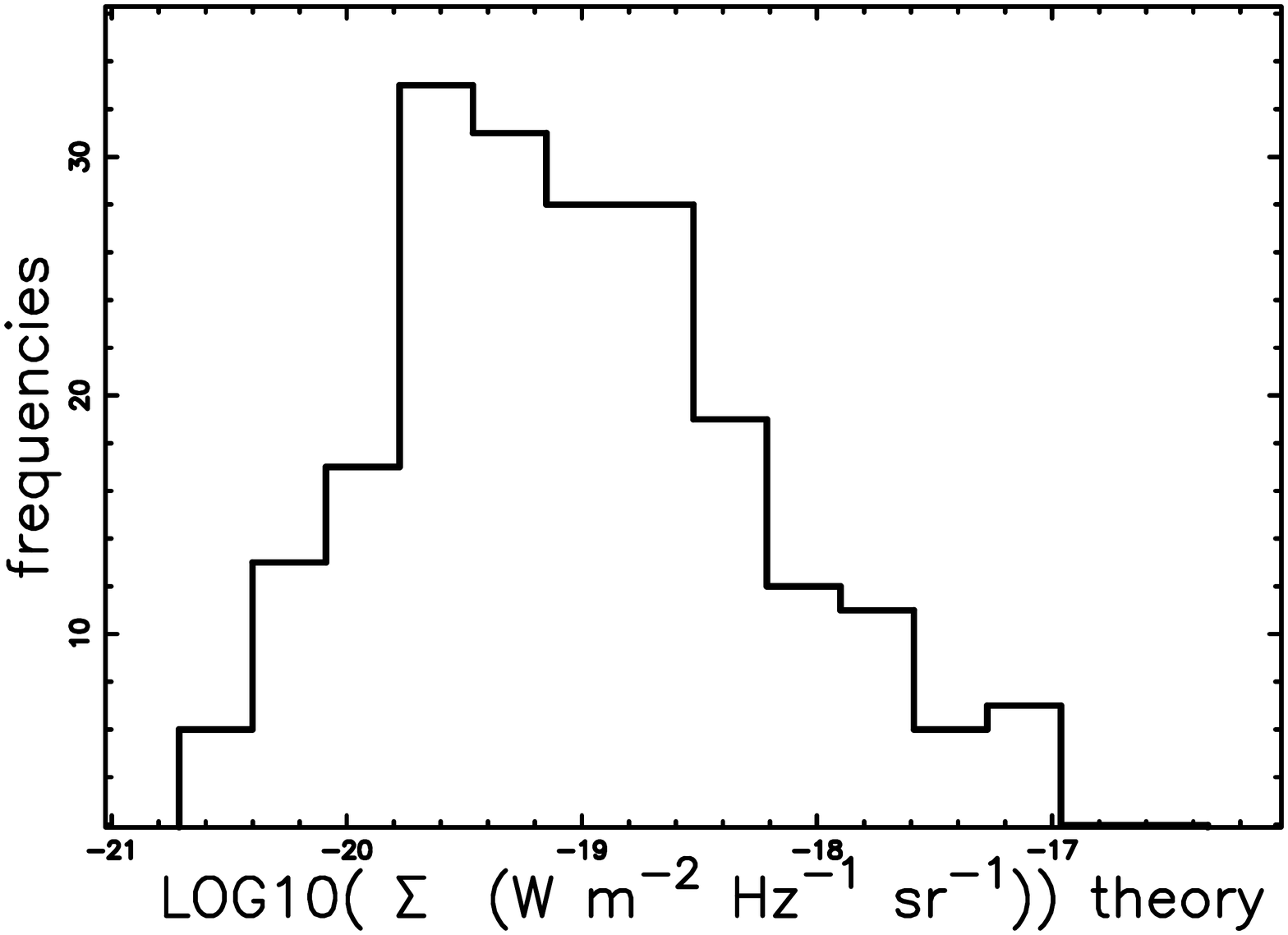}
\end {center}
\caption
{
Histogram  for the  \sigmad 
of our simulation in the framework of the inverse square model.
}
\label{simulation_green}
    \end{figure*}

The surface brightness
of  the Green's  catalog  
is reported  in Figure \ref{catalog_green}
and  our  simulation in 
Figure \ref{simulation_green};
see also  Table \ref{datagreen}
for a comparison between the data of the catalog 
and the simulation.

\begin{table}
\caption
{
Statistics of the observed \cite{Green2019}  
and simulated $\Sigma$ in the framework of the
inverse square model. 
}
 \label{datagreen}
 \[
 \begin{array}{lll}
 \hline
 \hline
 \noalign{\smallskip}
  parameter & observed   & theoretical \\
 \noalign{\smallskip}
 \hline
 \noalign{\smallskip}
\Sigma_{min} (pc)     & 1.42\,10^{-25}    &  1.27\,10^{-21}    \\
\overline{\Sigma}(pc) & 1.13\,10^{-18}    &  1.13\,10^{-18}  \\
\Sigma_{max} (pc)     & 4.34\,10^{-17}    &  6.17\,10^{-17}  \\
\noalign{\smallskip} 
\hline
 \end{array}
 \]
 \end {table}

\section{Extragalactic   Application}

\label{section_extragalactic}
\begin{figure*}
\begin{center}
\includegraphics[width=7cm]{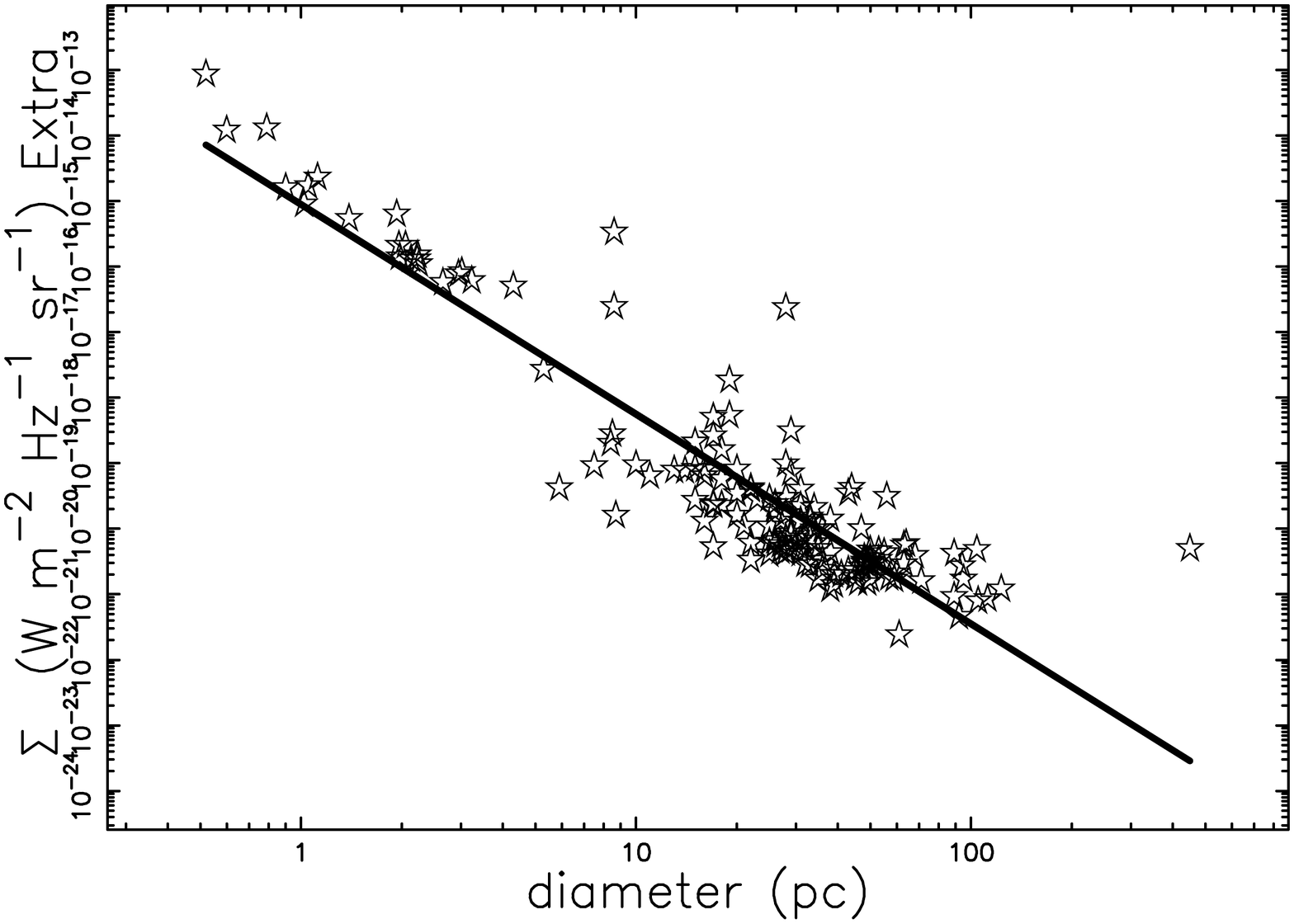}
\end {center}
\caption
{
Observed \sigmad relationship for the extragalactic case (empty stars)  
and  fitting  power law (full line)
with  parameters as in Table \ref{datafitsigmadextra}.
}
\label{obs_extra}
    \end{figure*}

A sample of   SNRs  in nearby galaxies  
with   diameter in pc and flux  
at 1.4 GHz in mJy  has been collected \cite{Urosevic2005}.
The connected  catalog is
available at \url{http://cdsweb.u-strasbg.fr/}.
The statistics of the \sigmad relationship for 
this extragalactic sample  
is reported  in Table \ref{datafitsigmadextra}
and  displayed in Figure \ref{obs_extra}.

\begin{table}
\caption
{
Statistics of the observed \sigmad extragalactic relationship
and the two parameters of the power law fit.
The theoretical parameters for the NCD case  are
$r_0=1\,pc$,
$r_{0,min}=2   \, r_0$,
$r_{0,max}=1.4 \, r_{0,min} $,
$v_0=4000\,km/s$,
$\delta=1.3$,
$d=1.1$,
$t_0=10\,yr$,   
$t_{min}= 2\,t_0$ 
and 
$t_{max}=2.9\,10^4$yr.
}
 \label{datafitsigmadextra}
 \[
 \begin{array}{lll}
 \hline
 \hline
 \noalign{\smallskip}
  parameter & observed   & theoretical \\
 \noalign{\smallskip}
 \hline
 \noalign{\smallskip}
D_{min} (pc)            & 0.51             & 4.05             \\
\overline{D}(pc)        & 34.6             & 34.23           \\
D_{max} (pc)            & 450              & 54.39           \\
\Sigma_{min} (pc)       & 2.4\,10^{-22}    &  1.87 \,10^{-17}    \\
\overline{\Sigma}(pc)   & 6.13\,10^{-16}   &  8.82\,10^{-16}  \\
\Sigma_{max} (pc)       & 8.6\,10^{-14}    &  4.14\,10^{-14}  \\
C                       & 8.85\,10^{-16}    & 2.64\,10^{-12}  \\
\alpha                  & -3.1              &  -2.96           \\
\noalign{\smallskip} 
\hline
 \end{array}
 \]
 \end {table}

\begin{figure*}
\begin{center}
\includegraphics[width=7cm]{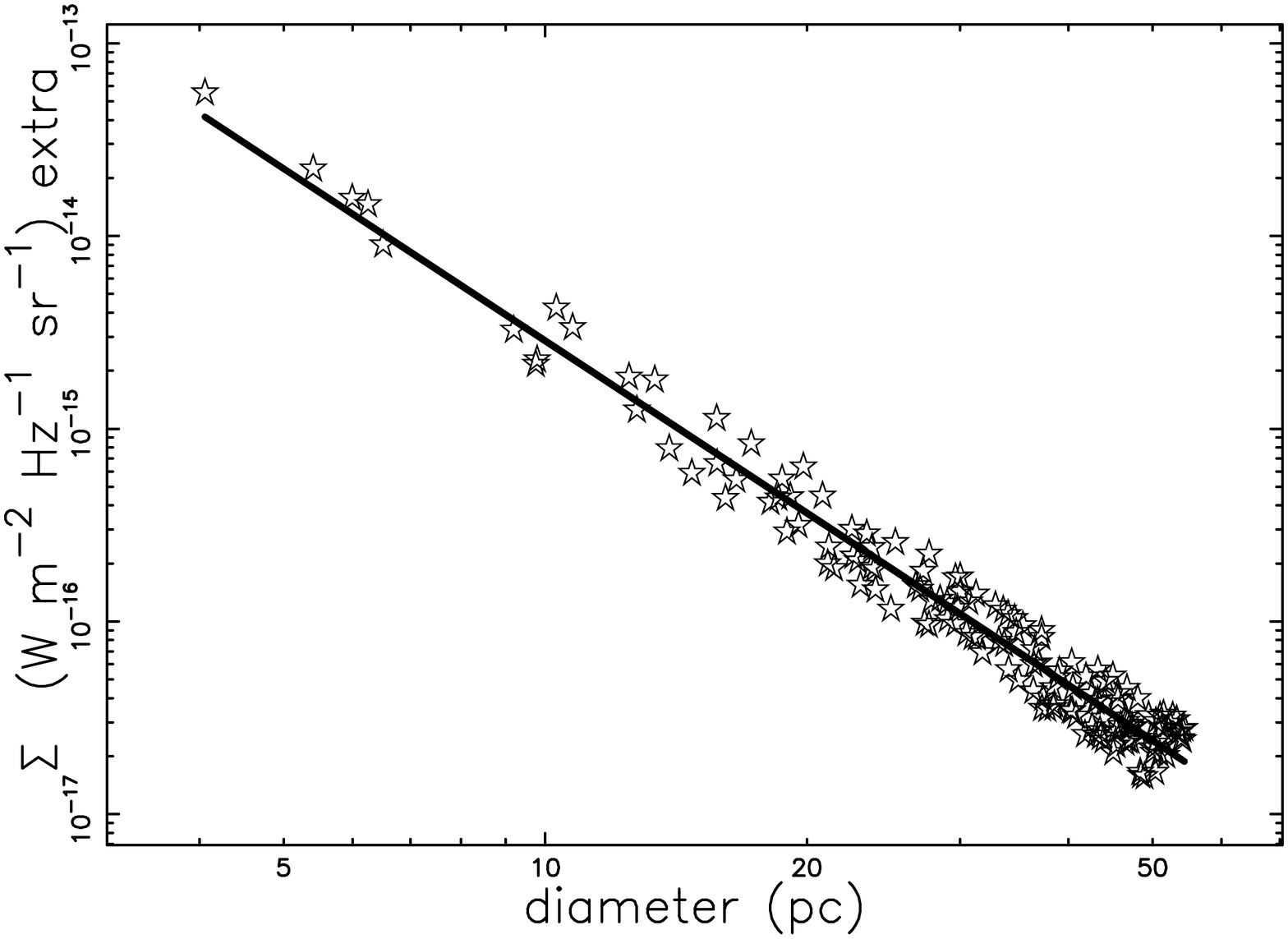}
\end {center}
\caption
{
Theoretical   \sigmad relationship for the extragalactic case 
in the framework of the NCD model
(empty stars)  
and  fitting  power law (full line)
with  parameters as in Table \ref{datafitsigmadextra}.
}
\label{extra_theo}
    \end{figure*}

The theoretical \sigmad is now  evaluated  
in the framework  of the NCD  model, see equation~(\ref{sigmadtncd}),
with a 
 numerical   procedure  which  is similar
to that  of the galactic  case but with the difference that there is  no
dependence of $r_0$ on $z$.
The numerical value of 
$r_0$ in the extragalactic case  is now randomly generated 
      according to the uniform distribution 
      between 
      a minimum value, $r_{0,min}$, 
      and
      a maximum value, $r_{0,max}$,
see Table  \ref{datafitsigmadextra}
and  Figure \ref{extra_theo}.

\section{Conclusions}

When an  analytical  law of motion is available, 
a theoretical 
\sigmad relationship
can be  derived as  a function of time.  
Here, in the framework of the energy conservation in the
thin-layer approximation, we have  derived 
one formula for  \sigmad  when the  density of the surrounding ISM 
scales as an inverse square   law, see equation
(\ref{sigmadtinversesquare}),
and  another formula for  the NCD case, see equation (\ref{sigmadtncd}).
The   two formulae allow simulating:
\begin {enumerate} 
\item
The galactic    \sigmad relationship  as  given 
by the data of \cite{Xu2005}, see  Figure \ref{sigmad_galactic_theory}
and Table  \ref{datafitsigmadgal}.
\item
The galactic    \sigmad relationship  as  given 
by a catalog of SNRs \cite{Green2019},
see Figure \ref{simulation_green} 
and 
Table \ref{datagreen}.
\item
An extragalactic  \sigmad  catalog \cite{Urosevic2005},
see  Figure \ref{extra_theo}  
and
Table  \ref{datafitsigmadextra}.
\end{enumerate}

\providecommand{\newblock}{}

\end{document}